\documentclass[12pt,preprint]{aastex}

\usepackage{natbib}

\shorttitle{Discovery of O-Ne-Mg-Rich Eejcta in Puppis~A} 
\shortauthors{Katsuda et al.}

\begin{document}

\title{Discovery of X-Ray--Emitting O-Ne-Mg-Rich Ejecta in the 
  Galactic Supernova Remnant Puppis~A}

\author{Satoru Katsuda\altaffilmark{1}, Una Hwang\altaffilmark{1, 2},
  Robert Petre\altaffilmark{1}, Sangwook Park\altaffilmark{3}, Koji
  Mori\altaffilmark{4}, and \\Hiroshi Tsunemi\altaffilmark{5}
}

\altaffiltext{1}{NASA Goddard Space Flight Center, Code 662, Greenbelt
        MD 20771}

\altaffiltext{2}{Department of Physics and Astronomy, The Johns Hopkins
        University, 3400 Charles Street, Baltimore, MD 21218}

\altaffiltext{3}{Department of Astronomy and Astrophysics, Pennsylvania
        State University, 525 Davey Laboratory, University Park, PA 16802}

\altaffiltext{4}{Department of Applied Physics, Faculty of Engineering,
University of Miyazaki, 1-1 Gakuen Kibana-dai Nishi, Miyazaki, 889-2192,
Japan}

\altaffiltext{5}{Department of Earth and Space Science, Graduate School
of Science, Osaka University, 1-1 Machikaneyama, Toyonaka, Osaka,
60-0043, Japan}

\begin{abstract}

We report on the discovery of X-ray--emitting O-Ne-Mg-rich ejecta in
the middle-aged Galactic O-rich supernova remnant Puppis~A with 
{\it Chandra} and {\it XMM-Newton}.  We use line ratios to identify a
low-ionization filament running parallel to the northeastern edge of
the remnant that requires supersolar abundances, particularly for O,
Ne, and Mg, which we interpret to be from O-Ne-Mg-rich ejecta.
Abundance ratios of Ne/O, Mg/O, and Fe/O are measured to be $\sim$2,
$\sim$2, and $<$0.3 times the solar values.  Our spatially-resolved
spectral analysis from the northeastern rim to the western rim
otherwise reveals sub-solar abundances consistent with those in the
interstellar medium.  The filament is coincident with several
optically emitting O-rich knots with high velocities.  If these are
physically related, the filament would be a peculiar fragment of
ejecta.  On the other hand, the morphology of the filament suggests
that it may trace ejecta heated by a shock reflected strongly off the 
dense ambient clouds near the northeastern rim.

\end{abstract}

\section{Introduction}

X-ray emission from evolved supernova remnants (SNRs) is dominated by 
the interstellar medium (ISM) swept-up by expanding SN ejecta.
Therefore, evolved SNRs had been considered to be more suitable for 
studies of high-Mach number ($> 10$) shock physics, ISM/shock
interactions, or the ISM itself, and less suitable for studies of
the SN explosion/nucleosynthesis.  However, over the last two decades,
the {\it ASCA}, {\it ROSAT}, {\it Chandra}, and {\it XMM-Newton} X-ray 
observatories have uncovered a number of ejecta features in
evolved SNRs such as the Vela SNR (e.g., Aschenbach 1995; Tsunemi et
al.\ 1999), the Cygnus Loop (e.g., Miyata et al.\ 1998; Katsuda et
al.\ 2008a), and several LMC (e.g., Hughes et al.\ 2003; Borkowski,
Hendrik, \& Reynolds 2006) and SMC SNRs (e.g., Park et al.\ 2003;
Hendrick, Reynolds, \& Borkowski 2005).

Puppis~A is one of the brightest SNRs in the X-ray sky and shows both
ISM/shock interactions and SN ejecta.  The large extent
($\sim$50$^{\prime}$ in diameter) as well as the high surface
brightness allow us to study detailed structures in the remnant.  It
has been suggested that the expanding shell of Puppis~A is interacting
with dense HI and CO clouds from the eastern (E) rim to the northern
(N) rim, based on its asymmetric X-ray surface brightness (e.g., Petre
et al.\ 1982) and the alignment of these dense clouds with the edge of the
remnant (e.g., Dubner \& Arnal 1988).  {\it Chandra} observations of
the most prominent cloud-shock interaction at the E rim can be
compared directly to scaled laboratory simulations to infer a mature
interaction of age 2000-4000 years (Hwang, Flanagan, \& Petre 2005;
Klein et al.\ 2003).  Thus, Puppis~A is an excellent astrophysical
laboratory where we can study detailed structures of cloud-shock
interactions.

On the other hand, evidence of SN ejecta has been detected in
Puppis~A.  From optical observations, Winkler \& Kirshner (1985)
discovered an O-rich fast-moving filament (named $\Omega$ filament) in
the northeastern (NE) portion of the remnant.  Subsequently, several 
O-rich optical knots have been found near the $\Omega$ filament, with
proper-motion vectors suggesting constant expansion from a common
center with a dynamical age of 3700$\pm$300\,yrs (Winkler et al.\
1988).  Recently, signatures of ejecta have been found in 
X-ray observations as well.  Hwang, Petre, and Flanagan (2008) noticed
Si-rich ejecta localized in the NE quadrant, based on a {\it Suzaku}
survey of this SNR [see also Hwang, Petre, \& Flanagan (2008) for a review
  of metal-rich ejecta indicated by earlier X-ray observations].
Katsuda et al.\ (2008b) also reported the discovery of fast-moving
metal-rich ejecta knots with blueshifted lines in the NE portion of
the remnant, based on {\it XMM-Newton} observations.  So far, the
composition of these ejecta features all indicate that Puppis~A
originated from a core-collapse SN.  This is consistent with the
presence of a central compact object (CCO) (Petre et al.\ 1996).

Here, we report the discovery of an O-Ne-Mg-rich ejecta filament in 
the NE quadrant of the remnant from {\it Chandra} and {\it XMM-Newton}
observations.  We also discuss the origin of the filament.

\section{Observations and Data Reduction}

\begin{figure}
\includegraphics[angle=0,scale=0.35]{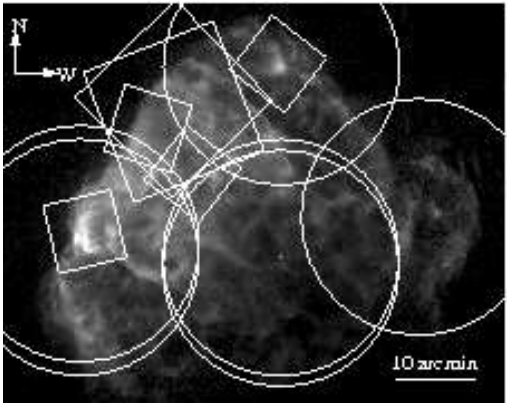}\hspace{1cm}
\caption{Fields of view of {\it XMM-Newton} (white circles) and {\it
  Chandra}  (white boxes) overlaid on a {\it ROSAT} HRI image of the
  entire Puppis~A SNR.  The data have been smoothed by Gaussian kernel
  of $\sigma = 15^{\prime\prime}$.  The intensity scale is square
  root.  
} 
\label{fig:hri}
\end{figure}

Puppis~A has been observed several times by the CCD cameras onboard
{\it Chandra} and {\it XMM-Newton}, with objectives such as 
the cloud-shock interactions, the CCO, or possible SN ejecta.
Together, these observations provide almost complete coverage of
this large X-ray remnant, as shown in Fig.~\ref{fig:hri}.

All the raw data from {\it XMM-Newton} were processed using version 
8.0.0 of the XMM Science Analysis Software (SAS).  We select X-ray
events corresponding to patterns
0--12\footnote{http://heasarc.nasa.gov/docs/xmm/abc/}.  We further
clean the data by rejecting high background (BG) intervals and
removing all the events in bad columns listed in Kirsch (2006).  After
the filtering, the data were vignetting-corrected using the SAS task
{\tt evigweight}.  As for {\it Chandra}, we reprocessed all the
level-1 event data, applying the standard data
reduction\footnote{http://cxc.harvard.edu/ciao/threads/createL2/} 
using CIAO ver.\ 4.1 and CALDB ver.\ 3.5.3.  There is no high BG
intervals during the {\it Chandra} observations.  These observations
are summarized in table~\ref{tab:obs}.

\section{Image Analysis}

The moderate spectral resolution of the CCD cameras together with good
spatial resolution of the X-ray telescopes onboard {\it Chandra} and
{\it XMM-Newton} allow us to investigate the detailed spectral
variation within Puppis~A.  Figure~\ref{fig:3col} shows an
exposure- and vignetting-corrected three-color image of merged {\it
  XMM-Newton} and {\it Chandra} data.  Red, green, and blue correspond
to 0.5--0.7\,keV (mostly O K-shell lines), 0.7--1.2\,keV (mostly Ne
K-shell lines), and 1.2--5.0\,keV bands, respectively.  We see clear
color variations such as a blue (hard emission) belt crossing
the remnant from the NE to the southwest (SW) as was previously
noticed with {\it ROSAT} and {\it ASCA} observations (Aschenbach 1993;
Tamura 1995; Hwang, Petre, \& Flanagan \ 2005; Hui \& Becker 2006), and
the western (W) limb that is enhanced in red (soft emission).  
We will briefly discuss their origins based on our spectral analyses
later in this paper.

Since the CCD spectra show strong lines especially from Ne as shown in
Fig.~\ref{fig:Ne_ratio} (a) which is a spatially integrated 
MOS1 spectrum, we generate a line ratio map of H-like Ly$\alpha$ line
to He$\alpha$ line blend for Ne in Fig.~\ref{fig:Ne_ratio} (b),
covering the same region as the three-color image
(Fig.~\ref{fig:3col}).  The energy bands used to generate the line 
ratio map are indicated in Fig.~\ref{fig:Ne_ratio} (a): 0.97--1.1\,keV
for H-like Ly$\alpha$ and 0.86--0.97\,keV for the He$\alpha$ blend.
We focus on a filamentary feature enclosed in the white box in
Fig.~\ref{fig:Ne_ratio} (b).  This feature (hereafter, the NE
filament) is clearly seen in Fig.~\ref{fig:3col} as a bright filament
with different color from those in its vicinity, indicating that it is
spectrally distinct.  In the NE filament, the surface brightness and
line ratio seem anti-correlated, with the low line ratio indicating
less emission from H-like ions.  Another feature showing such an
anti-correlation is marked by a black arrow in
Fig.~\ref{fig:Ne_ratio} (b).  This feature is the previously noted 
O-Ne-Mg-rich ejecta knot (Katsuda et al.\ 2008b) which is coincident
with the optical $\Omega$ filament (Winkler \& Kirshner 1985). The
spectral similarity between the NE filament and the $\Omega$ filament
leads us to speculate that the NE filament may also be an ejecta
feature.

\section{Spectral Analysis}

\subsection{Analysis of the NE Filament}

Figure~\ref{fig:fil} shows a close-up {\it Chandra} three-color image 
of the NE filament and its surroundings.  Based on the morphology, we 
divide the feature into four regions; two filaments and two knotty 
features.  These regions are marked by white ellipses labeled as
Fil-1/2 and Knot-1/2 in Fig.~\ref{fig:fil}.  The sizes of these
regions range from 20$^{\prime\prime}$ (the diameter for Knot-1's
circle) to 120$^{\prime\prime}$ (longer diameters for Fil-1/2's
ellipses), corresponding to $\sim$0.21--1.26\,pc at a distance of
2.2\,kpc (Reynoso et al.\ 2003).

\begin{figure}
\includegraphics[angle=0,scale=0.6]{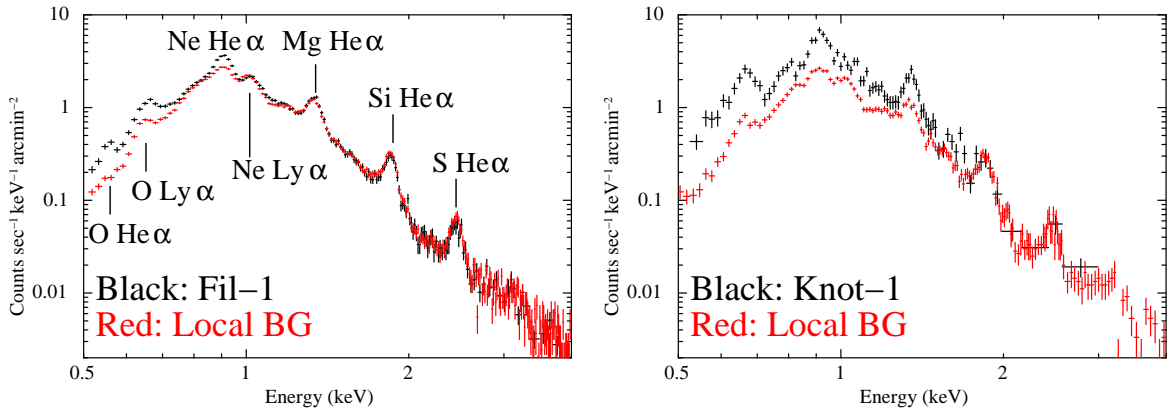}\hspace{1cm}
\caption{Left: Source spectrum from Fil-1 (black) with its
  background spectrum (red).  The names of strong lines are labeled. 
  Right: Same as left but for Knot-1.
}
\label{fig:spec_hikaku}
\end{figure}

Two {\it Chandra} observations (ObsIDs 5564 and 6371) cover Fil-1; we
extract spectra from each and combine them to improve signal-to-noise.
The other regions are covered by only one {\it Chandra} observation  
(ObsID 6371).  We subtract local BGs estimated from the surroundings
using the box and pie-shaped regions shown in Fig.~\ref{fig:fil}.  The
numbers of counts after the local BG subtraction are 7287, 6281, 2641,
and 11722 for Fil-1, 2, Knot-1, and 2, respectively.  To perform a 
$\chi^2$ test, each spectrum is grouped into bins with at least 20
counts.  Figure~\ref{fig:spec_hikaku} shows the Fil-1 and Knot-1
spectra in black together with their BG spectra in red.  The source
spectrum below $\sim$1.5\,keV shows excess emission which is dominated
by O and Ne K-shell lines for Fil-1, and O, Ne, and Mg K-shell lines
for Knot-1.

Based on the spectral similarity, we divide the four spectra into two 
groups: Fil-1/2 and Knot-1/2.  We simultaneously fit the two spectra
in each group (either Fil-1/2 or Knot-1/2).  We employ an absorbed,
single component, plane-parallel shock model [the {\tt Tbabs} (Wilms,
Allen, and McCray 2000) and the {\tt vpshock} model (NEI version 2.0)
(e.g., Borkowski, Lyerly, \& Reynolds 2001) in XSPEC v\,12.5.1].  We
freely vary the electron temperature, $kT_\mathrm{e}$, the ionization
timescale, $n_{\rm e}t$, and the volume emission measure (VEM $=\int 
n_\mathrm{e}n_\mathrm{H} dV$, where $n_\mathrm{e}$ and $n_\mathrm{H}$
are number densities of electrons and protons, respectively, and $V$
is the X-ray--emitting volume) for individual spectra.  Above, $n_{\rm
  e}t$ is the electron density times the elapsed time after shock
heating, and we take a range from zero up to a fitted maximum value.
The relative abundances of each group are tied together.  The
abundances of several elements, whose emission lines are prominent in
the spectrum, are treated as free parameters: O, Ne, Mg, Fe (=Ni) for
Fil-1/2, and O, Ne, Mg, Si (=S), 
Fe (=Ni) for Knot-1/2.  The other elements are fixed to the solar
values.  Throughout this paper, we employ solar abundances of Anders
\& Grevesse (1989).  We find that the abundances for Fil-1/2 are
super-solar ($>$1000 times the solar values).  In such a 
metal-rich plasma, it is difficult to precisely determine the absolute
abundances from the X-ray spectra since H does not contribute to the
continuum emission in this band.  We thus set the O abundance to be
2000 times the solar value as we did for our spectral analysis of the
$\Omega$ filament (Katsuda et al.\ 2008b; see also Winkler \& Kirshner
1985).  

We initially treated the hydrogen column density, $N_\mathrm{H}$, as a 
free parameter, and obtained the best-fit value of 
$\sim$5.5$\times$10$^{21}$\,cm$^{-2}$.  This value is quite large
compared with a typical value of 3$\times$10$^{21}$\,cm$^{-2}$ for
Puppis~A derived from recent {\it Suzaku} observations (Hwang, Petre,
\& Flanagan 2008) as well as our spectral analysis shown in the next
section.  Given that the color of Fil-1/2 is red in Fig.~\ref{fig:fil}
(which means that Fil-1/2 have soft spectra), the large absorption
would be strange.  Furthermore, it is not reasonable that only the NE
filament shows such a large absorption.  We thus fix $N_\mathrm{H}$ to
the typical value of 3$\times$10$^{21}$\,cm$^{-2}$ in the analysis of
Fil-1/2 and Knot-1/2.  Formally, the fit for Fil-1/2 is far from
acceptable ($\chi^2$~/~degrees of freedom = 480/248), but this is 
likely due to the imperfect subtraction of the local BGs.  While the
count rates after the local-BG subtraction are negative in some energy
bands, separate fits for Fil-1 and Fil-2 do not yield a better fit for
Fil-2 (while the abundances obtained for the two regions are
consistent with each other).  In this context, we introduce systematic
errors on the model, with the error for each spectral bin being
defined as $\sqrt{C_\mathrm{data}+(C_\mathrm{model}\times\alpha)}$,
where $C_\mathrm{data}$, $C_\mathrm{model}$, and $\alpha$ denote data  
counts, model counts, and the factor of the systematic error,
respectively.  We set the values of $\alpha$, so that we can obtain 
null hypothesis probabilities of the best-fit models to be $\sim$5\%. 

\begin{figure}
\includegraphics[angle=0,scale=0.6]{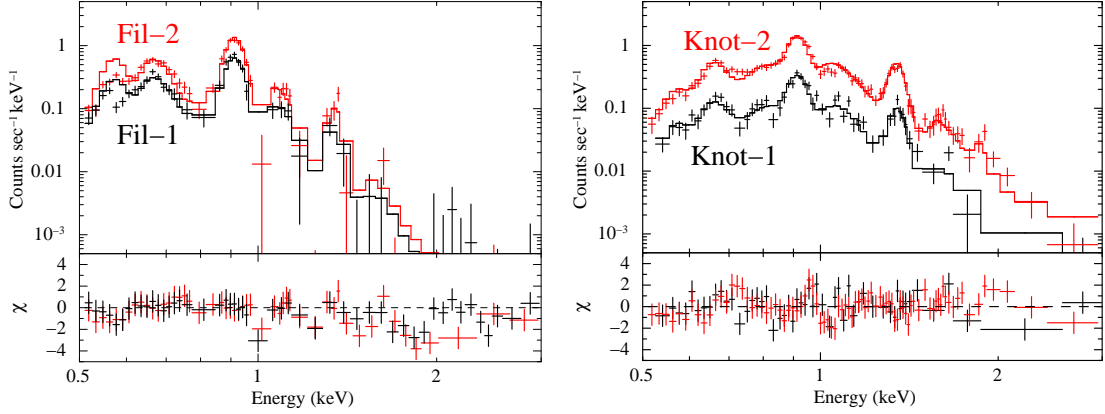}\hspace{1cm}
\caption{Left: Local-BG subtracted {\it Chandra} spectra
  of Fil-1/2 along with the best-fit models ($N_\mathrm{H}$ =
  3$\times$10$^{21}$\,cm$^{-2}$).  Lower panels show the 
  residuals. Right: Same as left but for Knot-1/2.
} 
\label{fig:fil_spec}
\end{figure}

Figure~\ref{fig:fil_spec} shows the local-BG subtracted spectra along
with the best-fit models.  The best-fit parameters and fit statistics
are listed in table~\ref{tab:fil}.  The elemental abundances are
measured relative to O as well as H.
The super-solar abundances of O, Ne, and
Mg and relative abundances of heavy elements such as Si and Fe in
Fil-1/2 suggest that the origin of the filament is O-Ne-Mg-rich
ejecta.  On the other hand, Knot-1/2 shows slight enhancements of O,
Ne, and Mg and sub-solar abundances of Si and Fe, suggesting that the
knots are mixtures of the O-Ne-Mg-rich ejecta and the ISM.  In
addition to the abundances, the best-fit electron temperatures are
also different between the two groups.  These spectral differences,
together with the different appearance between the two groups,
indicate that they could be located at different positions along the
line of sight.  We note that the low temperatures and ionization
timescales found in Fil-1/2 result in the low line ratios of H-like to
He-like ions seen in Fig.~\ref{fig:Ne_ratio} (b).

\subsection{Radially-Resolved Spectral Analysis}

\begin{figure}
\includegraphics[angle=0,scale=0.8]{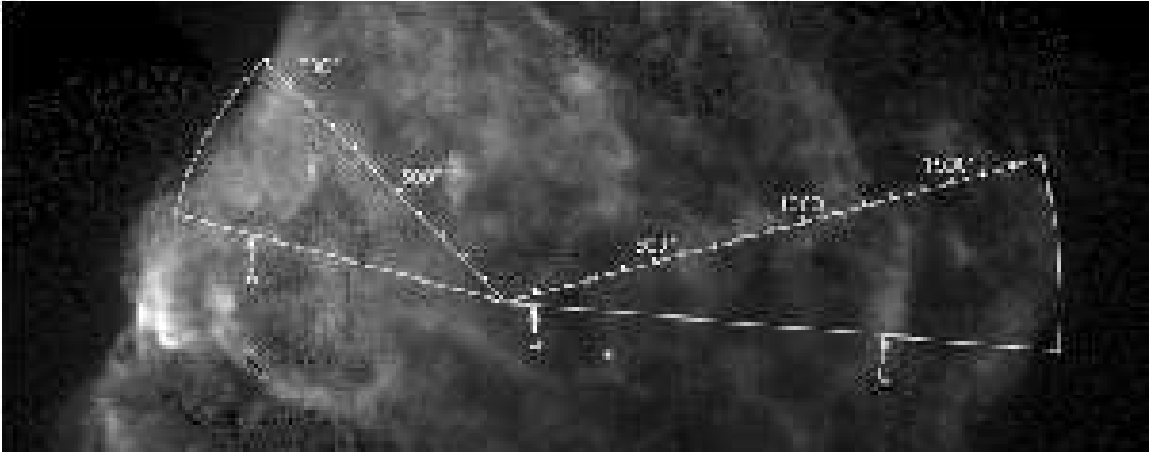}\hspace{1cm}
\caption{The area used for radially-resolved spectral analysis
  overlaid on an X-ray wide band (0.5--5.0\,keV) image from the merged
  {\it XMM-Newton} and {\it Chandra} data.  The SNR center, [(RA, DEC) =
  (125.6145, -42.9579) J2000], is the expansion center of O-rich
  optical fast-moving knots (Winkler et al.\ 1988).  We divide this
  area into thin annular regions for our spectral analysis.  Arrows
  with letters A, B, and C indicate the radial positions where we show
  example spectra in Fig.~\ref{fig:ex_spec}.
}
\label{fig:area}
\end{figure}

We have performed spectral analyses of the NE filament including both
the filamentary and knotty features, and found them to be 
rich in O, Ne, and Mg.  If other ejecta regions are found, then it is
possible that the contact discontinuity and the reverse shock might be
identified within the remnant based on the extent of the
X-ray--emitting ejecta.  To search for these features, we perform
radially-resolved spectral analysis of pie-shaped areas from the NE
rim to the W rim shown in Fig.~\ref{fig:area}.  The center of the
pie-shaped area is the inferred expansion center of the O-rich optical
knots (Winkler et al.\ 1988).  We divide the area into thin annular
regions such that each region contains $\sim$10000 photons.  The
thickness of these regions ranges from 3$^{\prime\prime}$ to
11$^{\prime\prime}$ for {\it Chandra} covering part of the NE portion,
and 13$^{\prime\prime}$ to 180$^{\prime\prime}$ for {\it XMM-Newton}
covering the entire area.  As BG, we use regions outside Puppis~A in
the same observations for {\it Chandra}.  As for {\it XMM-Newton}, we
use the data set accumulated from blank sky observations prepared by
Read \& Ponman (2003) because it provides much better statistics than
the BG available from Puppis~A observations.  We have checked for a
few regions that different BGs (either the blank sky or the source
free region around Puppis~A) do not affect the spectral parameters
within 90\% confidence limits.

\begin{figure}
\includegraphics[angle=0,scale=0.5]{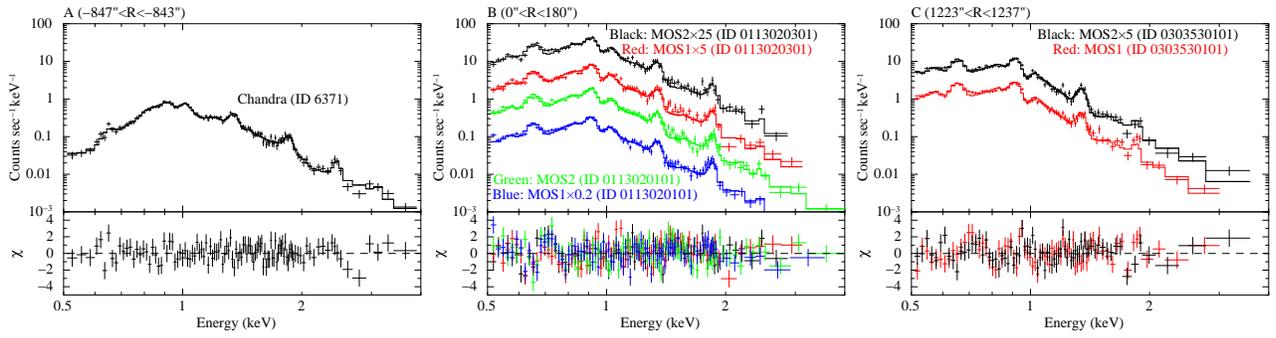}\hspace{1cm}
\caption{Example spectra from three regions indicated as arrows in 
  Fig.~\ref{fig:area} along with the best-fit models.  Lower
  panels show the residuals.
} 
\label{fig:ex_spec}
\end{figure}

We fit the spectra using an absorbed, single-component {\tt vpshock} 
model.  Free parameters are $N_\mathrm{H}$, $kT_\mathrm{e}$, 
$n_{\rm e}t$, VEM, and abundances of O, Ne, Mg, Si, S, and Fe (=Ni).
Abundances of other elements are fixed to the solar values.  For some
regions in the W rim, the source emission above 2\,keV is so weak
that we tie the abundance of S to that of Si.  Figure~\ref{fig:ex_spec} 
depicts example spectra from three regions whose radial positions
are indicated by arrows with letters A, B, and C in
Fig.~\ref{fig:area}.  The fit levels for these regions are typical, 
with reduced $\chi^2$-values ranging from 1.02 (A) to 1.56 (C).
These examples show good matches between the data and the model.  The
best-fit parameters for all regions are plotted as a function of
radius in Fig.~\ref{fig:rad_res}.  The black and red data are derived
from {\it XMM-Newton} and {\it Chandra}, respectively.  Note that in
these plots we exclude the region of the NE filament
($-$810$^{\prime\prime}$$<$R$<$$-$720$^{\prime\prime}$), since we 
already know that at least two spectral components (i.e., the O-Ne-Mg-rich
ejecta and its surroundings) are required to reproduce spectra there.
As we can see in the Fig.~\ref{fig:rad_res} bottom right panel (i.e., 
reduced $\chi^2$), this model represents all the spectra fairly well.  
The general consistency of the results between {\it XMM-Newton}
and {\it Chandra} validates our analysis.  

The fitted abundances are all lower than the solar values.  The
abundances relative to O are consistent with the solar ratios within a
factor of 2, showing that there is no strong evidence of ejecta in
this region except for the O-Ne-Mg-rich NE filament.  The electron
temperature is almost constant at 0.7\,keV in the E portion of the
SNR, while it gradually decreases toward the W rim to $\sim$0.4\,keV.
This temperature decrease presumably makes the W limb appear reddish
in the three-color image of the SNR (see, Fig.~\ref{fig:3col} left). 
The gradual decrease of the ionization timescale from the E rim to the
W rim may reflect the gradual increase of electron density toward E
(Petre et al.\ 1982).  The slight increase of $N_\mathrm{H}$ toward
the NE rim might reflect the extinction by the dense cloud sitting in
front of Puppis~A (Reynoso et al.\ 1995).   With the exception of the
NE rim, the value of $N_\mathrm{H}$ is almost constant at 
3$\times$10$^{21}$\,cm$^{-2}$, consistent with recent {\it Suzaku}
observations (Hwang, Petre, \& Flanagan 2008).  Given that the area we 
investigate lies both inside and outside the hard emission (blue 
color in Fig.~\ref{fig:3col} left) belt across Puppis~A (e.g.,
Aschenbach 1993; see section~3), the blue belt-like feature is more
likely caused by variations in the temperature and the ionization
timescale than in column density.

\section{Discussion and Summary}

The X-ray surface brightness map and the Ne K line ratio map reveal a 
distinct filamentary feature (i.e., the NE filament) and two knots in
the NE portion of Puppis~A.  Our spectral analysis shows 
that these features are rich in O, Ne, and Mg, with abundance ratios
of Ne/O, Mg/O twice solar and that of Fe/O less than solar.  The
super-solar ($>$1000) fitted metal abundances of the NE filament
clearly show that its origin is likely O-Ne-Mg-rich ejecta.  The
relative abundances of the O-Ne-Mg-rich ejecta can then be used to
estimate the progenitor mass.  We compare the relative abundances
measured in the filament with those predicted by the theoretical
calculations (Rauscher et al.\ 2002) for different progenitor masses
of 15\,M$_\odot$, 25\,M$_\odot$, 30\,M$_\odot$, and 35\,M$_\odot$.  We 
find that the data agree with those calculations for the O-Ne-Mg-rich
layers in progenitor stars of $\lesssim$ 25\,M$_\odot$ masses.  We note
that the same conclusion can be derived by using the 
relative abundances of the $\Omega$ filament (Katsuda et al.\ 2008b),
since the relative abundances of Fil-1/2 are consistent with those of
the $\Omega$ filament.  This is also consistent with the
progenitor-mass range of 15\,M$_\odot$--25\,M$_\odot$ inferred from
comparisons of relative abundances of metal-rich (especially Si-rich)
ejecta in the NE portion with several nucleosynthesis models (Hwang,
Petre, \& Flanagan 2008).  We caution however that we only measured
abundances of just a small fraction of the total ejecta, and so
the estimate of the progenitor mass is still not conclusive.

To try to identify the contact discontinuity and the reverse shock, we
performed radially-resolved spectral analysis from the NE rim 
(including the O-Ne-Mg-rich NE filament) to the W rim.  The spectral
analysis revealed that all the spectra except for the NE filament show
sub-solar metal abundances ($\sim$0.5 times the solar values) whose
relative abundances are consistent with the solar values within a
factor of 2---consistent with those from {\it Suzaku} observations
covering the same region (Hwang, Petre, \& Flanagan 2008).  This
result suggests that the X-ray emission is dominated by the ISM
{\it everywhere} except for the NE filament.  Interpretations of the
results could be (1) the outer ejecta heated by the reverse shock have
already cooled, while the inner ejecta have not yet been heated by the
reverse shock, (2) metal-rich ejecta could not be detected (except for
the NE filament) because of their faintness compared with the ISM, or
(3) most of the ejecta has not yet been heated by the reverse shock,
although this last case seems unlikely for a remnant of several
thousand years' age such as Puppis A.  With the current data, this
remains an open question.

\begin{figure}
\includegraphics[angle=0,scale=0.4]{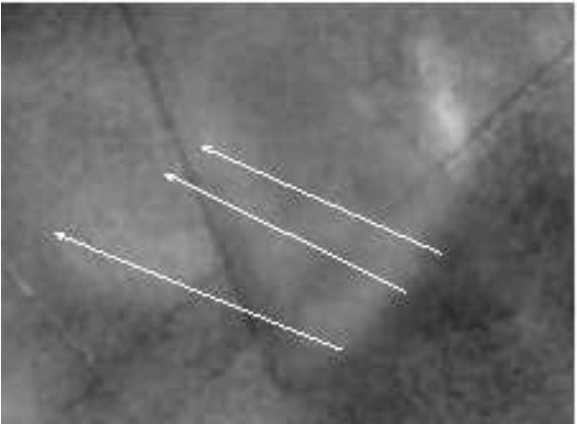}\hspace{1cm}
\caption{X-ray wide band (0.5--5\,keV) image focused on the 
  NE filament.  Proper motion vectors for 1000\,yr of three optical
  O-rich fast-moving knots are indicated as arrows (Winkler et al.\
  1988; updated by Garber et al.\ 2010).
} 
\label{fig:X_opt}
\end{figure}

In the area we investigated, the only region that shows evidence
of SN ejecta is the small region where the O-Ne-Mg-rich NE filament
and knots are located.  This result may indicate that they are peculiar
fragments of ejecta inside and/or outside the SNR.  We then
note that the NE filament is coincident with three\footnote{In the 
  vicinity of these knots, recent optical observations identified 
  some more knots whose proper motion vectors are similar with each 
  other (Garber et al.\ 2010).} optical O-rich fast-moving knots 
(Winkler et al.\ 1988).  Figure~\ref{fig:X_opt} shows the proper
motion vectors of the O-rich knots on the X-ray image, in which we see
that all the vectors, $\sim$0$^{\prime\prime}$.18\,yr$^{-1}$
($\sim$1900\,km\,sec$^{-1}$ at a distance of 2.2\,kpc) toward the NE,
are similar to each other.  If the NE filament is physically
associated with these optical knots, we expect the same proper motions
for the NE filament as well.

\begin{figure}
\includegraphics[angle=0,scale=0.35]{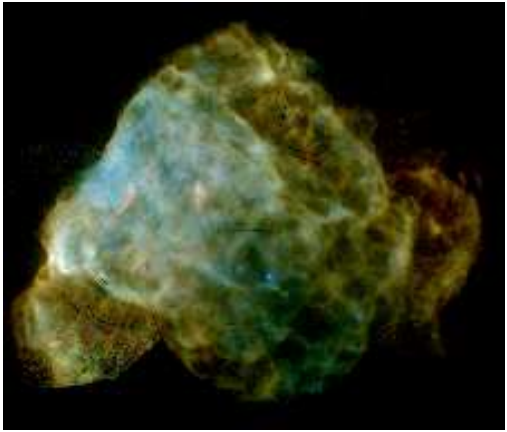}\hspace{1cm}
\caption{Three-color image of the merged {\it XMM-Newton} and
  {\it Chandra} data.  Red, green, and blue correspond to
  0.5--0.7\,keV (mostly O K-shell lines), 0.7--1.2\,keV (mostly Ne
  K-shell lines), and 1.2--5.0\,keV bands, respectively.  The data
  have been smoothed by Gaussian kernel of $\sigma =
  4^{\prime\prime}$.  The intensity scale is logarithmic.  
} 
\label{fig:3col}
\end{figure}

\begin{figure}
\plottwo{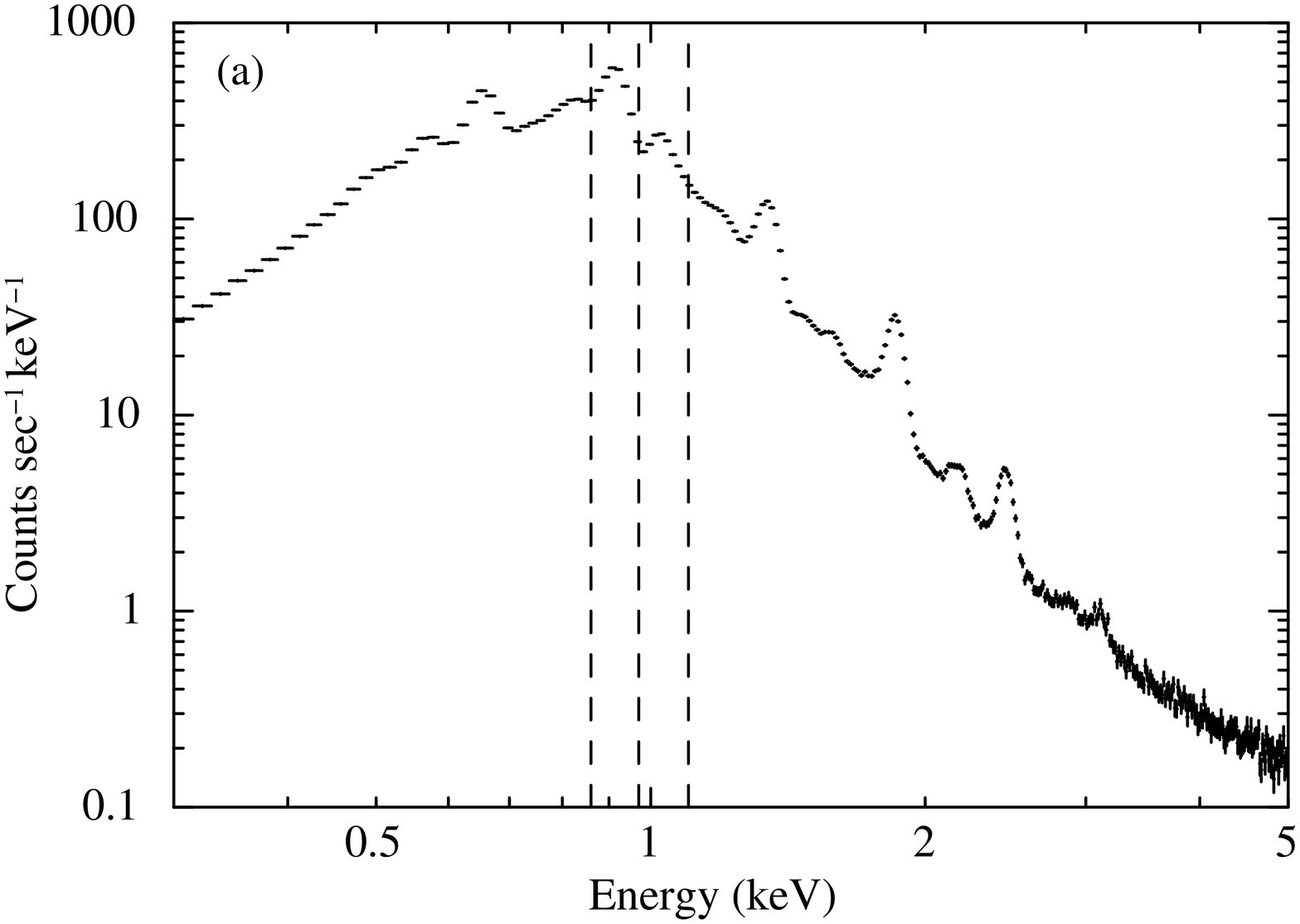}{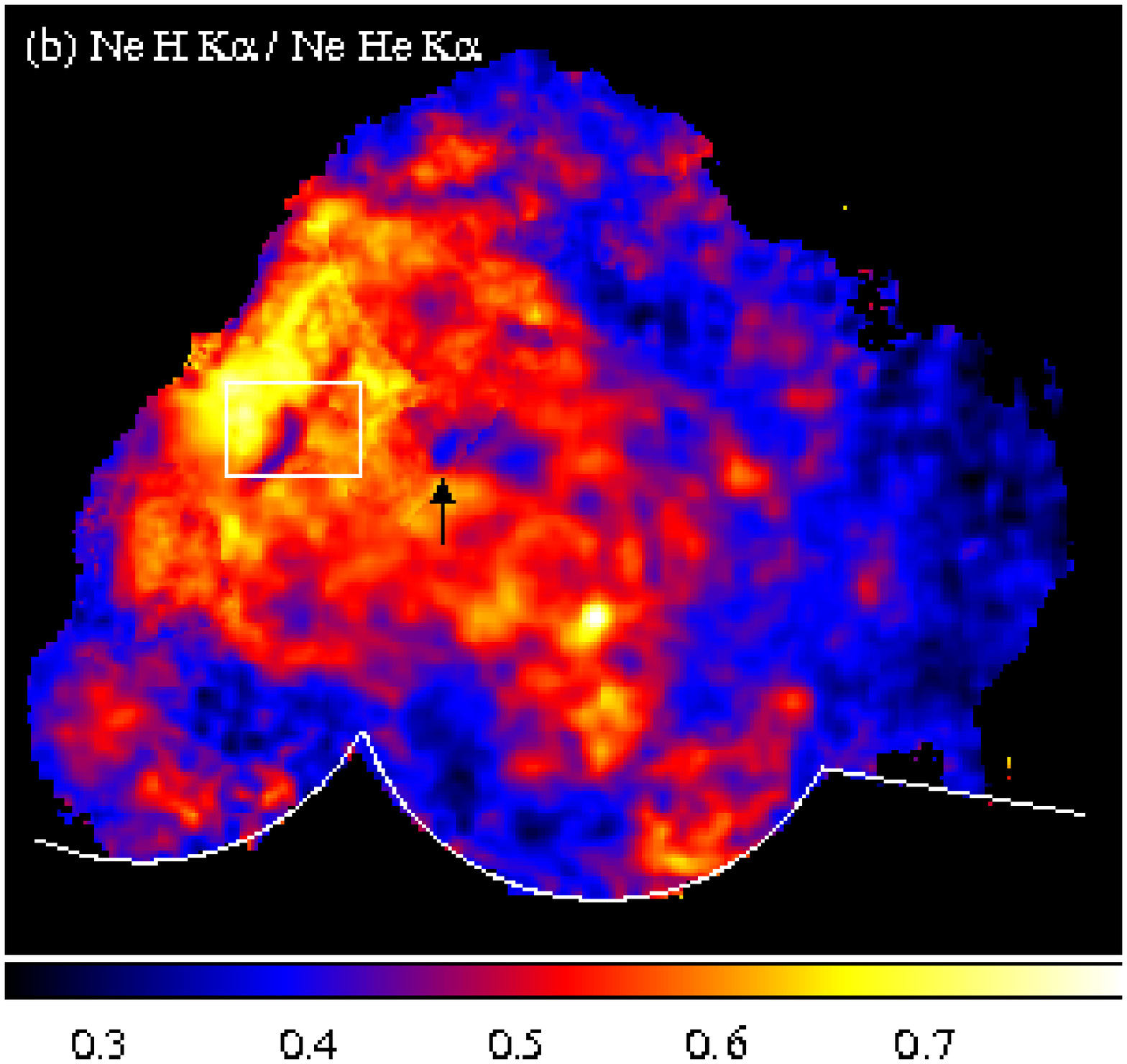}
\caption{(a): MOS1 spectrum integrated by all {\it XMM-Newton}
  observations.  The energy bands used for generating the line-ratio
  map (Fig.~\ref{fig:Ne_ratio}b) are indicated as dashed lines.
  (b):  Line-ratio map of H-like Ly$\alpha$ to He$\alpha$ of Ne,
  covering the same area of the three-color image shown in
  Fig.~\ref{fig:3col}.  The white line at the bottom of the figure
  marks the edge of fields of view. 
} 
\label{fig:Ne_ratio}
\end{figure}

\begin{figure}
\includegraphics[angle=0,scale=0.7]{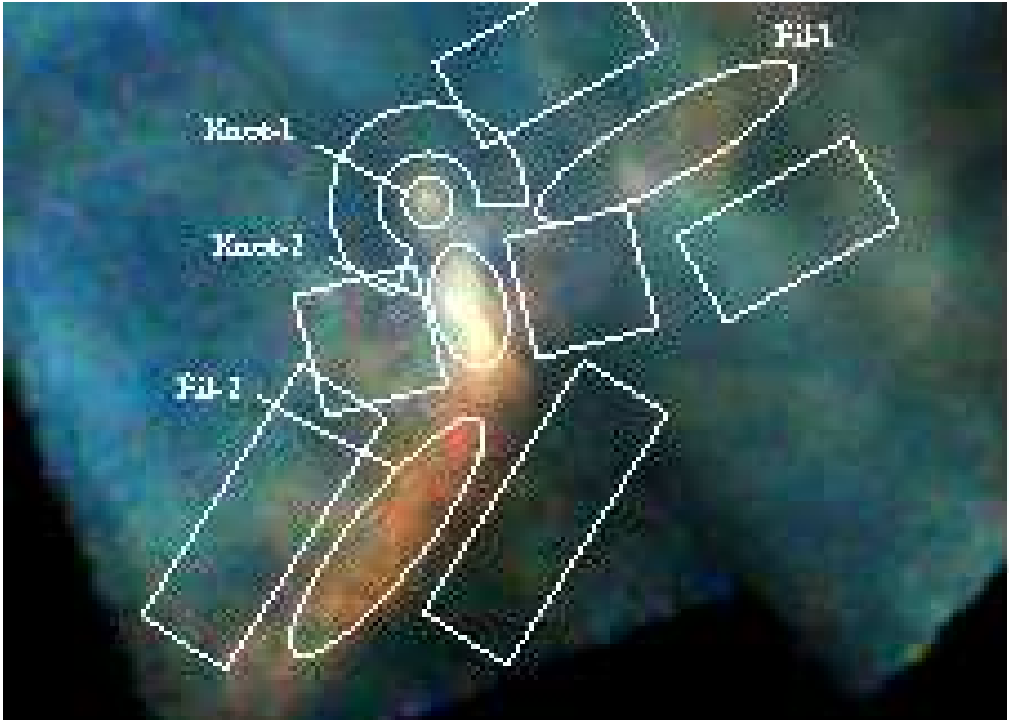}\hspace{1cm}
\caption{Expanded three-color {\it Chandra} image focused on the white
  box region in Fig.~\ref{fig:Ne_ratio}b.  White ellipse/circle labeled as
  Knot-1/2 and Fil-1/2 are spectral extraction regions.  The local
  backgrounds are estimated from the box or pie-shaped regions close
  to the source regions.  BG spectra from two box regions are merged. 
}
\label{fig:fil}
\end{figure}

\begin{figure}
\includegraphics[angle=0,scale=0.6]{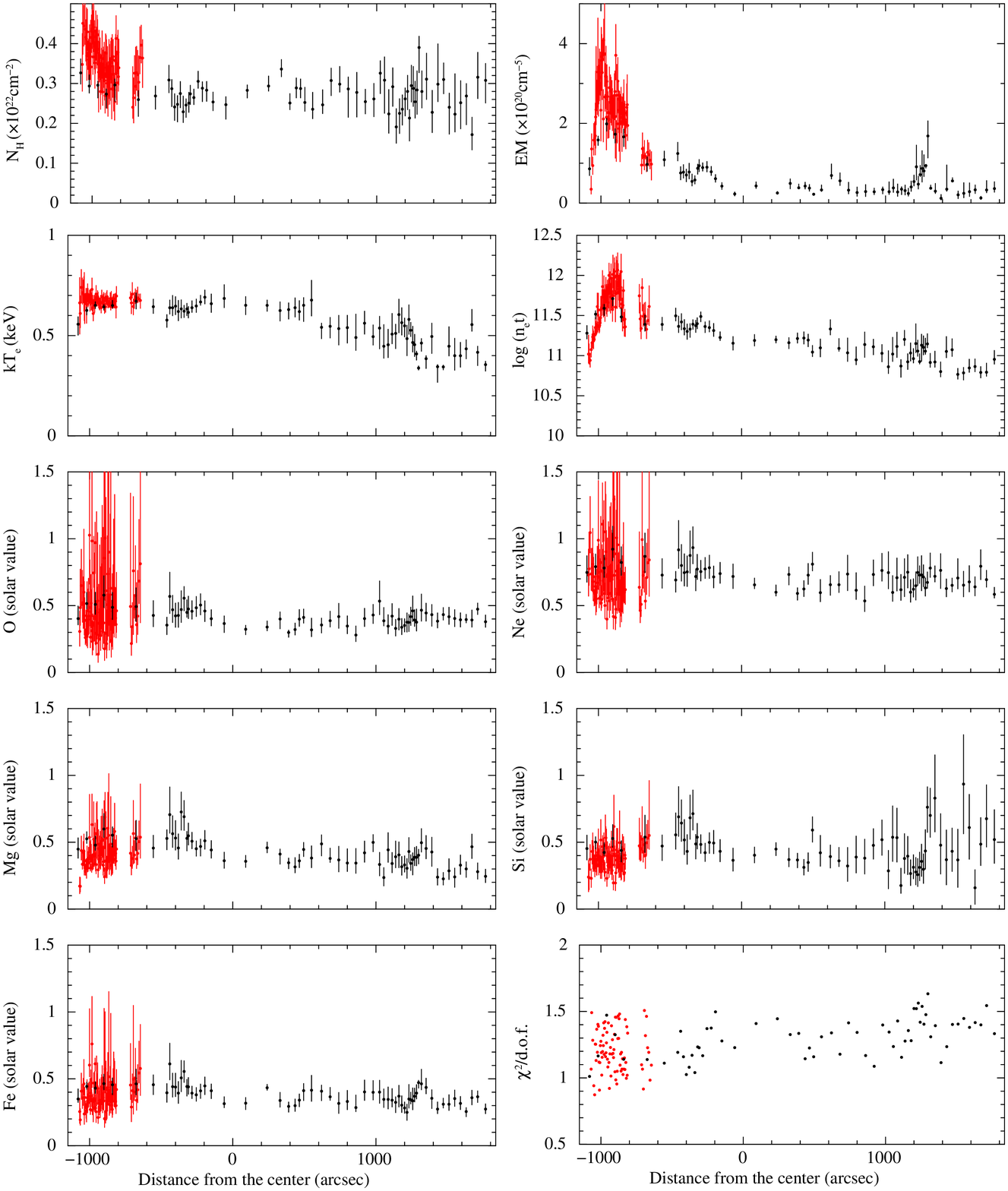}\hspace{1cm}
\caption{Results from radially-resolved spectral analysis as a
  function of the distance from the SNR center.  Left is to the NE,
  and right is to the W.  Red and black data
  correspond to {\it Chandra} and {\it XMM-Newton} data,
  respectively.  The results in the region including the NE filament 
  (-810$^{\prime\prime}<$R$<$-720$^{\prime\prime}$) are not well
  represented by the model employed and are excluded in the plots. 
} 
\label{fig:rad_res}
\end{figure}

On the other hand, it is interesting to note that the NE filament runs 
parallel to the NE edge of the remnant (see, Figs.~\ref{fig:3col} and 
\ref{fig:Ne_ratio} (b)).  This suggests that the morphology of the
filament is related to the NE edge of the SNR where the forward shock
is suggested to be interacting with dense clouds (e.g., Reynoso et
al.\ 1995).  Then, we can speculate that the filament traces ejecta
heated by either a strong reflected shock off the dense ambient clouds
in the NE rim or a reverse shock strongly developed in this portion of
the SNR.  Given that we see evidence of ejecta only in the NE
filament, we may interpret the inner side of the filament to be the
reflection shock and the outer side to be a boundary between the
O-Ne-Mg-rich ejecta and either the ISM or ejecta rich in lighter
elements.  A problem for this scenario is that the expected
temperature signature at the reflected shock is not seen
(Fig.~\ref{fig:rad_res}); the temperature should be higher in the 
twice-shocked outer regions than in the singly shocked inner regions
(e.g., Hester et al.\ 1994).  It is also uncomfortable that the
filament spans only a portion of the NE edge, whereas it appears that
the entire NE edge is interacting with dense clouds (e.g., Reynoso et
al.\ 1995).  However, these signature might be less clear due to
projection effects and/or possible density inhomogeneities of the
ejecta; the denser ejecta will tend to have the lower temperature, as
the temperature is partly determined by the condition that the ram
pressure, $\rho v^2$, be approximately constant.  One way to test the
reflection shock scenario would be to measure the proper motion of the
filament, as it should then be slower than the freely expanding
ejecta.

We tried measuring proper motions of the NE filament, based on X-ray
images of {\it Einstein} in 1979, {\it ROSAT} in 1992, 1993, and 1994,
and {\it Chandra} in 2006.  Although we indeed found a possible
motion, it is difficult to derive conclusive results, because there is
no point-like source in the {\it Einstein} and {\it ROSAT} images to
be used for image registrations, and the brightness of the NE filament
seems to change during these observations.  In this respect, future
{\it Chandra} observations of the NE filament are desired to firmly
detect the possible proper motion and brightness change, and to
conclude the origin of the NE filament.

\acknowledgments

We are grateful to J.\ Garber and collaborators for fruitful 
discussions about the optical O-rich knots.  We thank K.\ Borkowski for
discussions about Fe L line emissivities.  S.K.\ is supported by a  
JSPS Research Fellowship for Young Scientists and in part by NASA
grant NNG06EO90A.

\clearpage

\begin{deluxetable}{cccccc}
\tabletypesize{\scriptsize}

\tablecaption{Summary of {\it XMM-Newton} and {\it Chandra} observations}
\tablewidth{0pt}
\tablehead{
\colhead{Field}&\colhead{Obs.ID}&\colhead{Instruments}
&\colhead{Obs.Date}&\colhead{Coordinate (J2000)} & \colhead{Exposure
  Time (ksec)} 
}
\startdata
\multicolumn{6}{c}{{\it XMM-Newton}}\\
Center1 &0113020101 & MOS  & 2001-04-15 & 125.486,  -43.0272& 22.0\\
Center2 &0113020301 & MOS  & 2001-11-08 & 125.486, -43.0053& 10.8\\
North &0150150101 & MOS+pn & 2003-04-17 & 125.487, -42.6161 & 6.3\\
East1 &0150150201 & MOS+pn & 2003-06-25 & 126.045, -42.9778 & 3.6\\
East2 &0150150301 & MOS+pn & 2003-06-25 & 126.046, -43.0058 & 4.3\\
West &0303530101 & MOS+pn & 2005-10-09 & 125.100, -42.9203& 9.8\\
\hline
\multicolumn{6}{c}{{\it Chandra}}\\
North & 1949 & ACIS-S & 2002-03-09 & 125.529, -42.6426 & 19.9\\
Northeast1&1950 & ACIS-S & 2001-10-01 & 125.841, -42.7031 & 14.9\\
East &1951 & ACIS-S& 2001-11-04 & 126.048, -42.9141 & 11.7\\
Northeast2 & 5564& ACIS-I & 2005-09-04 & 125.782, -42.6906 & 34.5\\
Northeast3 &6371& ACIS-I & 2006-02-11 & 125.788, -42.6981 & 28.3\\

\enddata
\tablecomments{The fields of view of MOS and pn are
  30$^\prime$-diameter circles.  Those of ACIS-S and ACIS-I are
  8.5$^\prime$-side and 17.1$^\prime$-side boxes, respectively.}

\label{tab:obs}
\end{deluxetable}

\begin{deluxetable}{lcccc}
\tabletypesize{\scriptsize}
\tablecaption{Spectral fit parameters} 
\tablewidth{0pt}
\tablehead{
\colhead{Parameter}
&\colhead{Fil-1}&\colhead{Fil-2}&\colhead{Knot-1}&\colhead{Knot-2}
}
\startdata
$N_\mathrm{H}$
($\times$10$^{21}$\,cm$^{-2}$)\dotfill&\multicolumn{4}{c}{3 (fixed)}\\
$kT_{\rm e}$ (keV)\dotfill&  0.43 (0.28--0.71)& 0.33 (0.26--0.54) 
& 0.79 (0.55--1.26) &0.80 (0.60--1.06) \\
log($n_{\rm e}t$/cm$^{-3}$\,sec)\dotfill& 10.20 (9.82--10.74) & 10.24
(9.88--10.61) &  10.45 (10.25--10.68) & 10.59 (10.44--10.82)\\  
O (O$_\odot$)\dotfill& \multicolumn{2}{c}{2000 (fixed)} & 
\multicolumn{2}{c}{1.8 (0.9--3.2)} \\ 
Ne (Ne$_\odot$)\dotfill& \multicolumn{2}{c}{3400 (2800--4500)} &
\multicolumn{2}{c}{3.7 (1.9--9.8)} \\
Mg (Mg$_\odot$)\dotfill& \multicolumn{2}{c}{3400 (1800--8000)} &
\multicolumn{2}{c}{3.5 (1.8--6.5))} \\ 
Si (Si$_\odot$)\dotfill& \multicolumn{2}{c}{1 (fixed)} &
\multicolumn{2}{c}{0.8 (0.3--2.0)} \\
Fe (Fe$_\odot$)\dotfill& \multicolumn{2}{c}{0 ($<$470)} &
\multicolumn{2}{c}{0.9 (0.4--2.5)} \\
Ne/O (Ne$_\odot$/O$_\odot$)\dotfill& \multicolumn{2}{c}{1.75
  (1.45--2.15)} & \multicolumn{2}{c}{2.05 (1.93--2.18)} \\  
Mg/O (Mg$_\odot$/O$_\odot$)\dotfill& \multicolumn{2}{c}{1.85
  (1.30--2.65)} & \multicolumn{2}{c}{2.00 (1.84--2.17)} \\  
Si/O (Si$_\odot$/O$_\odot$)\dotfill& \multicolumn{2}{c}{---} &
\multicolumn{2}{c}{0.50 (0.15--0.85)} \\   
Fe/O (Fe$_\odot$/O$_\odot$)\dotfill& \multicolumn{2}{c}{0
  ($<$0.25)} & \multicolumn{2}{c}{0.49 (0.41--0.58)} \\ 
$\int n_\mathrm{e}n_\mathrm{H} dl^{\mathrm a}$
($\times$10$^{17}$\,cm$^{-5}$)\dotfill& 0.21 (0.14--0.35)& 
0.62 (0.39--1.39)&360 (170--900)& 420 (200--880) \\ 
$\int n_\mathrm{e}n_\mathrm{O} dl^{\mathrm a}$
($\times$10$^{17}$\,cm$^{-5}$)\dotfill& 0.36 (0.24--0.60)& 
1.06 (0.66--2.37)&0.55 (0.26--1.38)& 0.64 (0.31--1.35) \\ 
$\chi^2$/d.o.f.\dotfill&\multicolumn{2}{c}{287/250}&\multicolumn{2}{c}{204/186}\\ 
\enddata
\tablecomments{$^{\mathrm a}$VEM normalized by the region area; 
  $dl$ is the plasma depth and $n_\mathrm{O}$ is the number density of
  O.  Values in parentheses indicate the 90\%
  confidence ranges (calculated by including the systematic errors of 
  40\% and 10\% to the data for Fil-1/2 and Knot-1/2, respectively). 
  The errors of relative abundances are calculated after fixing
  $N_\mathrm{H}$, $kT_{\rm e}$, and log($n_{\rm e}t$) at the best-fit
  values.  Other metals are fixed to the solar values (Anders \&
  Grevesse 1989).}
\label{tab:fil}
\end{deluxetable}

\end{document}